\documentclass[12pt]{article}

\catcode`\@=11
\global\arraycolsep=2pt
\usepackage{amsbsy,amssymb,latexsym,amsfonts,amsmath}
\usepackage{graphicx,color}

\begin{document}

\begin{flushright}
RUP-21-14
\parbox{4.2cm}

\end{flushright}

\vspace*{0.7cm}

\begin{center}
{ \Large Recursive structure of the Gau\ss{} hypergeometric function and boundary/crosscap conformal block}
\vspace*{1.5cm}\\
{Yu Nakayama}
\end{center}
\vspace*{1.0cm}
\begin{center}

Department of Physics, Rikkyo University, Toshima, Tokyo 171-8501, Japan

\vspace{3.8cm}
\end{center}

\begin{abstract}
The Gau\ss{}  hypergeometric function shows a recursive structure which resembles those found in conformal blocks. We compare it with the recursive structure of the conformal block in boundary/crosscap conformal field theories that is obtained from the representation theory. We find that the pole structure perfectly agrees but the recursive structure in the Gau\ss{} hypergeometric function is slightly ``off-shell". 
\end{abstract}

\thispagestyle{empty} 

\setcounter{page}{0}

\newpage

\section{Introduction}
In our studies of conformal field theories, evaluating conformal blocks is of utmost importance for theoretical purposes as well as for practical purposes.  General conformal blocks, however, do not admit easily accessible analytic expressions. Even when they do, we still appreciate a direct way to evaluate their numerical values in order to perform practical computations such as conformal bootstrap analysis (see e.g. \cite{Poland:2018epd}  for a review).  This is because we need precision  to do the numerical conformal bootstrap, which might be lacking, e.g. if we use the analytic expression such as an integral representation if any. 

In \cite{Kos:2013tga}\cite{Kos:2014bka}, they have developed a recursive method to evaluate conformal blocks for four-point functions. In two-dimensions, a similar technique to compute Virasoro conformal block was first introduced by Zamolodchikov \cite{Zamolodchikov:1984eqp}, which remains the only practical method to compute it (except for special cases) even today \cite{Perlmutter:2015iya}. In higher dimensions, the recursion relation becomes the most efficient way to compute the conformal block numerically as well as analytically. General ideas to extend the recursive methods to compute conformal blocks in conformal field theories with conformal defects are in further investigations \cite{Lauria:2017wav}\cite{Kobayashi:2018okw}\cite{Lauria:2018klo}. The origin of the recursion relation can be traced back to a representation theory of conformal algebra and the structure of their matrix elements as a function of the conformal dimensions \cite{Penedones:2015aga}\cite{Erramilli:2019njx}\cite{Poland:2021xjs}.

On the other hand, the conformal blocks are related to generalized hypergeometric functions \cite{Isachenkov:2016gim}\cite{Chen:2016bxc}\cite{Schomerus:2016epl}\cite{Fortin:2016dlj}\cite{Karateev:2017jgd}\cite{Kravchuk:2017dzd}\cite{Isachenkov:2018pef}\cite{Chen:2019gka}\cite{Li:2019dix}\cite{Comeau:2019xco}\cite{Li:2019cwm}\cite{Buric:2020zea}\cite{Buric:2021ywo}\cite{Buric:2021ttm}. They can be mapped to eigenfunctions of underlying integrable models. Understanding the recursive structure in terms of these special functions may shed further light on the origin of these special functions as well as mathematical understanding of conformal blocks.

This small note aims to investigate the relation between the recursive structure of the hypergeometric functions and the conformal blocks in the simplest case of the Gau\ss{} hypergeometric function and the boundary/crosscap conformal block. We see that they are similar but slightly different in the sense that the former is ``off-shell". We, nevertheless, show that the pole structures perfectly agree with each other.

\section{Recursion relation}
The Gau\ss{} hypergeometric function\footnote{In this paper, the Pochhammer symbol is defined by $(a)_n = \frac{\Gamma(a+n)}{\Gamma(n)} = a (a+1) \cdots (a+n-1)$.}
\begin{align}
{}_2F_1(a,b;c;z) = \sum_{n \ge 0} \frac{(a)_n (b)_n}{(c)_n} \frac{z^n}{n!}
\end{align}
shows simple poles when the third argument is a non-negative integer (i.e. $-c = m \in \mathbb{Z}_{\ge 0}$). The direct computation shows that the residues themselves are the Gau\ss{} hypergeometric functions with shifted argument \cite{NIST}
\begin{align}
\lim_{c\to -m } \frac{{}_2F_1(a,b;c;z)}{\Gamma(c)} = \frac{(a)_{m+1}(b)_{m+1}}{(m+1)!}z^{m+1}{}_2F_1(a+m+1,b+m+1;m+2;z) \ , 
\end{align}
where
\begin{align}
\Gamma(c) \sim \frac{(-1)^m}{(c+m)m!}    
\end{align}
as $-c \to m \in \mathbb{Z}_{\ge 0}$.

If we regard the Gau\ss{} hypergeometric function as a meromorphic function with respect to the third argument $c$, we can reproduce it by knowing all the residues at poles on the complex $c$ plane. This idea, in turn, leads to  the recursion relation for the Gau\ss{} hypergeometric function:
\begin{align}
&{}_2 F_1 (a,b;c;z)  = \cr
&1 + \sum_{m \ge 0}\frac{(-1)^m}{(c+m) m!}\frac{(a)_{m+1}(b)_{m+1}}{(m+1)!}z^{m+1}{}_2F_1(a+m+1,b+m+1;m+2;z) \ . \label{hrecursion} 
\end{align}
Here we have supplied the remainder function $1$ by noting $\lim_{c \to \infty; a,b, \text{fixed}}{}_2 F_1 (a,b;c;z) = 1$. If we know the Gau\ss{} hypergeometric function up to order $z^N$, then we can evaluate the coefficient of order $z^{N+1}$ from the right hand side of \eqref{hrecursion}. During the explicit computation, we observe that this recursion relation essentially gives the partial fractional decomposition of $\frac{1}{(c)_m}$ that appears at order $z^m$.

Now let us move on to the application to conformal field theories. It is well-known that conformal block for a conformal defect with co-dimension one is described by the Gau\ss{} hypergeometric function \cite{McAvity:1995zd}\cite{Liendo:2012hy}\cite{Nakayama:2016cim}\cite{Nakayama:2016xvw}\cite{Gadde:2016fbj}\cite{Giombi:2020xah}. To set the stage, we will focus on conformal field theories on a real projective plane, but we will make a comment on the boundary conformal field theories when appropriate.

Consider a conformal field theory on a $d$ dimensional real projective plane defined by the identification $\vec{x} \to -\frac{\vec{x}}{\vec{x}^2}$ of the Euclidean space. The two-point function of scalar primary operators $O_1$ with conformal dimension $\Delta_1$ and $O_2$ with conformal dimension $\Delta_2$ can be parameterized by
\begin{align}
\langle O_1(\vec{x}_1) O_2(\vec{x}_2) \rangle = \frac{(1+\vec{x}_1^2)^{\frac{\Delta_{21}}{2}} (1+\vec{x}_2^2)^{\frac{\Delta_{12}}{2}}}{(\vec{x}_1-\vec{x}_2)^{{\Delta_1 + \Delta_2}}} G(\eta) \ , 
\end{align}
where the crosscap cross-ratio is defined by $\eta = \frac{(\vec{x}_1-\vec{x}_2)^2}{(1+\vec{x}_1^2)(1+\vec{x}_2^2)}$. We have introduced the notation $\Delta_{12} = \Delta_1 - \Delta_2$ and $\Delta_{21} = \Delta_2 - \Delta_1$. 

By using the operator product expansion of $O_1 \times O_2 = \sum_i c_{12i} O_i$, the two-point function can be decomposed into a sum of conformal blocks
\begin{align}
G(\eta) = \sum_i c_{12i} A_i \eta^{\frac{\Delta_i}{2}} {}_2F_1(\frac{\Delta_{12} +\Delta_i} {2},\frac{\Delta_{21}+\Delta_i}{2};\Delta_i + 1-\frac{d}{2};\eta) \ ,
\end{align}
where $A_i$ is the one-point function of $O_i$ on the projective plane.
The conformal block was derived in various ways. The direct way is to evaluate the operator product expansions of $O_1$ and $O_2$ from the explicit formula in \cite{McAvity:1995zd} and evaluate the resultant one-point functions. An alternative way is to solve the conformal Casimir equation that the two-point functions must satisfy, which gives the hypergeometric differential equation, with the boundary condition that is compatible with the operator product expansions.
See e.g. \cite{Hasegawa:2016piv}\cite{Hasegawa:2018yqg} for applications in critical systems.

If we are interested in boundary conformal field theories rather than conformal field theories on real projective space, the conformal block is obtained by ``analytic continuation" of $\eta \to -\frac{\xi}{4}$.  See e.g. \cite{Giombi:2020xah}.

With the explicit form in terms of the Gau\ss{} hypergeometric function,  we may use the recursion relation to evaluate the boundary/crosscap conformal block (with the factor $\eta^{\frac{\Delta}{2}}$ removed):
\begin{align}
h(\Delta_{12},\Delta,\eta)  = {}_2F_1(\frac{\Delta_{12} +\Delta} {2},\frac{\Delta_{21}+\Delta}{2};\Delta + 1-\frac{d}{2};\eta) \ . 
\end{align}

When we substitute \eqref{hrecursion}, we obtain the ``off-shell recursion relation":
\begin{align}
h(\Delta_{12},\Delta,\eta) = 
1 + \sum_{m \ge 0}\frac{(-1)^m}{(\Delta+ 1 -\frac{d}{2}+m)}\frac{(\frac{\Delta_{12} + \Delta}{2})_{m+1}(\frac{\Delta_{21} + \Delta}{2})_{m+1}}{m! (m+1)!} \times \cr \times \eta^{m+1} {}_2F_1(\frac{\Delta_{12} + \Delta  + 2m + 2 }{2},\frac{\Delta_{21} + \Delta + 2m + 2}{2};m+2;z) \ .
\end{align}
This ``recursion relation" is exact, yet it is ``off-shell" in the sense that the residues that appear on the right hand side are not the conformal blocks but are the Gau\ss{} hypergeometric function with general arguments. 

It would look more like a recursion relation of the Zamolodchikov type if we could replace $\Delta$ with $\Delta^{N_A}_* = \frac{d}{2} -N_A$ and $\Delta_*^{N_A,s} = \frac{d}{2} + N_A$ (with $m = N_A-1$) in the residues as
\begin{align}
    h(\Delta_{12},\Delta,\eta) \sim 1 + \sum_{N_A \ge 1}\frac{(-1)^{N_A-1}}{(\Delta- \Delta^{N_A}_*)}\frac{(\frac{\Delta_{12} + \Delta^{N_A}_*}{2})_{N_A}(\frac{\Delta_{21} + \Delta^{N_A}_*}{2})_{N_A}}{(N_A-1)! (N_A)!} \cr \times \eta^{N_A}h(\Delta_{12},\Delta_*^{N_A,s}, \eta) \  \label{rec?} 
\end{align}
so that $\Delta$ appears only in the denominator.
Here, $\sim$ means that they agree only near each pole. In other words, we cannot recursively construct the conformal block from the information of the conformal block of the lower order alone (from this particular method).  
We could attribute the difference to the remainder function $h(\Delta_{12},\Delta \to \infty, \eta)$ (rather than $1$ used in \eqref{hrecursion}), but this limit of the Gau\ss{} hypergeometric function is non-trivial and gives an exponentially large (i.e. $e^{\frac{\Delta}{4}\eta}$ for small positive $\eta$) contribution as first studied by Watson \cite{Watson} more than a century ago, who used the integral representation.

On the other hand, we can use \eqref{rec?} to study the pole structure of the conformal block independently predicted from the representation theory. For general conformal blocks, when the dimension $\Delta$ of the exchanged operator corresponds to a short representation (i.e. it has a specific conformal dimension $\Delta_*$ and the particular descendant with conformal dimension $\Delta_*^s$ becomes a primary state that is also null), it develops a pole:
\begin{align}
h(\Delta) \sim \frac{R}{\Delta -\Delta_*} h(\Delta_*^s)  \ ,
\end{align}
where the residue is given by the conformal block of the null state. When we know all the poles $\Delta_i$ in the complex plane then the meromorphic function $h(\Delta)$ can be reconstructed, giving rise to a recursion relation. 

The location of the poles, or the condition for the null vectors, has been worked out in the literature (see e.g.  \cite{Penedones:2015aga}).
In our case, we recall that only the null vectors of type III with spin zero appear in the boundary/crosscap conformal block. More explicitly they are located at $\Delta_*^{N_A} = \frac{d}{2} - N_A$ and $\Delta_* ^{N_A,s}= \frac{d}{2}+ N_A$ so that the null operator is $(P^2)^{N_A} O_{\Delta_*}$. 

The coefficient $R$ in the residue consists of three parts $R = M^L Q M^R$, where $M^R = \langle \alpha| \phi_1 \phi_2 \rangle$, $Q = \langle \alpha | \alpha \rangle$, $M^L = \langle B| \alpha \rangle$, where $|\alpha \rangle$ is the descendant states (that becomes null when $\Delta \to \Delta_*$).  They are explicitly computed \cite{Lauria:2017wav} in boundary conformal field theories from the general methods presented in \cite{Penedones:2015aga}:
\begin{align}
Q_{\mathrm{III}} &= -\frac{1}{(-16)^{N_A}(N_A-1)!N_A!(\frac{d}{2}-N_A-1)_{2N_A}} \frac{(\frac{d}{2}-N_A-1)}{(\frac{d}{2}+N_A-1)}   \cr
M_{\mathrm{III}}^R  &= 4^{N_A} \left(\frac{\Delta_{12} + \Delta_*^{N_A}}{2}\right)_{N_A}  \left(\frac{\Delta_{21} + \Delta_*^{N_A}}{2}\right)_{N_A} \cr
M_{\mathrm{III}}^L & = (-4)^{N_A}\left(\frac{\frac{d}{2}-N_A+1}{2} \right)_{N_A} \left(\frac{\frac{d}{2}-N_A}{2} \right)_{N_A}
\end{align}
Multiplying these three factors, we can confirm the proposed residues from our ``off-shell recursion relation" \eqref{rec?}:
\begin{align} M_{\mathrm{III}}^L  Q_{\mathrm{III}}  M^R_{\mathrm{III}}  = 
 -(4)^{-N_A}\frac{(\frac{\Delta_{12} + \Delta_*}{2})_{N_A}(\frac{\Delta_{21} + \Delta_*}{2})_{N_A}}{(N_A-1)! (N_A)!} 
\end{align}
Note that we should do an analytic continuation of $\eta \to -\frac{\xi}{4}$  to convert our formulas for the real projective plane to the boundary conformal field theories, and this explains an extra factor of $(-4)^{N_A}$.

\section{Discussions}

As we have alluded in the introduction, not only the simplest boundary/crosscap conformal block but more complicated conformal blocks can be represented by generalized hypergeometric functions. They have their own recursive structure and it would be interesting to investigate them and compare them from the viewpoint of the representation theory of conformal symmetries.  
 
For example, let us take a look at four-point conformal block evaluated on a $z=\bar{z}$ limit \cite{El-Showk:2012cjh}. It is known that the scalar conformal block of identical scalars is expressed by
\begin{align}
G_{\Delta}(z) = \left(\frac{z^2}{1-z}\right)^{\frac{\Delta}{2}} {}_3 F_2 \left(\frac{\Delta}{2},\frac{\Delta}{2},\frac{\Delta}{2}- \frac{d}{2}+1;\frac{\Delta+1}{2},\Delta-\frac{d}{2}+1;\frac{z^2}{4(z-1)} \right) \ . 
\end{align}
We can see that it develops a pole when the fourth and fifth argument become non-positive integers. 
The formula \eqref{hrecursion}  has an obvious generalization:
\begin{align}
{}_3 F_2 (\alpha,\beta,\gamma;\delta,\epsilon ;x)  = \cr
1 + \sum_{m \ge 0}\frac{(-1)^m}{(\delta+m) m!}\frac{(\alpha)_{m+1}(\beta)_{m+1} (\gamma)_{m+1}}{(\epsilon)_{m+1}(m+1)!} \times \cr
\times x^{m+1}  {}_3F_2(\alpha +m+1,\beta+m+1,\gamma+m+1;m+2,\epsilon+m+1;x) 
\label{hrecursion3}
\end{align} 
and similar formula applies to the fifth argument.
This again gives the ``off-shell" recursion relation for the scalar four-point conformal block, which resembles the one proposed in \cite{Kos:2013tga}\cite{Kos:2014bka}. This is ``off-shell" in the sense that the right hand side cannot be written as a special case of  conformal blocks.
Nevertheless, we can  use it to verify the pole structures as well as the residues of the conformal block proposed in  \cite{Kos:2013tga}\cite{Kos:2014bka}. Here we have two series of poles, one from $\delta = \frac{\Delta+1}{2} = - m$ and the other from $\epsilon = \Delta-\frac{d}{2}+1 = -m$. They  correspond to type I and type III null vectors \cite{Penedones:2015aga}.

From the holographic viewpoint, the conformal block has a geometric Witten diagram expression \cite{Hijano:2015zsa}. In the case of the four-point functions, they are given by
\begin{align}
G = &\frac{4\Gamma(\Delta)^2}{\Gamma(\frac{\Delta+\Delta_{12}}{2})\Gamma(\frac{\Delta-\Delta_{12}}{2})\Gamma(\frac{\Delta+\Delta_{34}}{2})\Gamma(\frac{\Delta-\Delta_{34}}{2})} \cr
&\times \int_{\gamma_{12}}\int_{\gamma_{34}} G_{b\partial}(y(\lambda),x_1) G_{b\partial}(y(\lambda),x_2) G_{bb}(y(\lambda),y(\lambda');\Delta) G_{b\partial}(y(\lambda'),x_3) G_{b\partial}(y(\lambda'),x_4)  \ ,
\end{align}
where the integration is over the geodesic $\gamma_{12}$ and $\gamma_{34}$ connecting two boundary points.  Bulk-boundary propagator and bulk-bulk propagator are defined by
\begin{align}
G_{b\partial}  (y,x_i) & = \left(\frac{u}{u^2 + |x - x_i|^2} \right)^\Delta_i \cr 
G_{bb}(y,y';\Delta) &= e^{-\Delta\sigma(y,y')}{}_2F_1 (\Delta,\frac{d}{2};\Delta+1-\frac{d}{2};e^{-2\sigma(y,y')})
\end{align}
in terms of the geodesic distance $\sigma (y,y')$.

In this expression, the origin of the pole structure is separated into two pieces. The one from the third argument of the Gau\ss{} hypergeometric function in the bulk-bulk propagator, which explains type III null vectors. The other comes from $\Gamma(\Delta)$ in the overall coefficient and one may say that it is not directly related to the AdS dynamics. By itself, it only gives the off-shell recursive structure, so it would be more interesting if we could give a holographic meaning to the recursion relation.

Finding the residue coefficient $R$ in the recursion relation of the Zamolodchikov type for conformal blocks is a non-trivial task. In the case of the Virasoro conformal block, $R$ proposed in \cite{Zamolodchikov:1984eqp} was confirmed by using higher equations motion in the boundary Liouville conformal field theory \cite{Zamolodchikov:2003yb} (see also \cite{Bertoldi:2004yk} for its proof in the classical regime). It would be interesting to pursue similar directions.

\section*{Acknowledgements}
A part of this note was originally prepared for intensive lecture courses on conformal field theories in higher dimensions at Osaka University and Tokyo Institute of Technology. This work is in part supported by JSPS KAKENHI Grant Number 21K03581.

\end{document}